\documentclass[%
reprint,
superscriptaddress,
amsmath,
amssymb,
aps,
prl,
floatfix,
]{revtex4-2}

\usepackage{graphicx}
\usepackage{dcolumn}
\usepackage{bm}
\usepackage{hyperref}

\usepackage[utf8]{inputenc}
\usepackage[T1]{fontenc}
\usepackage{mathptmx}
\usepackage{gensymb}

\DeclareSymbolFont{epsilon}{OML}{cmm}{m}{it}
\DeclareMathSymbol{\epsilon}{\mathord}{epsilon}{"0F}

\DeclareUnicodeCharacter{2212}{-}

\begin{document}

\preprint{APS/123-QED}

\title{Protein folding as a jamming transition}

\author{Alex T. Grigas}
\affiliation{Graduate Program in Computational Biology and Bioinformatics, Yale University, New Haven, Connecticut, 06520, USA}
\affiliation{Integrated Graduate Program in Physical and Engineering Biology, Yale University, New Haven, Connecticut, 06520, USA}
\author{Zhuoyi Liu}
\affiliation{Department of Mechanical Engineering and Materials Science, Yale University, New Haven, Connecticut, 06520, USA}
\affiliation{Integrated Graduate Program in Physical and Engineering Biology, Yale University, New Haven, Connecticut, 06520, USA}
\author{Jack A. Logan}
\affiliation{Department of Mechanical Engineering and Materials Science, Yale University, New Haven, Connecticut, 06520, USA}
\author{Mark D. Shattuck}
\affiliation{Benjamin Levich Institute and Physics Department,
The City College of New York, New York, New York 10031, USA}
\author{Corey S. O'Hern}
\affiliation{Department of Mechanical Engineering and Materials Science, Yale University, New Haven, Connecticut, 06520, USA}
\affiliation{Graduate Program in Computational Biology and Bioinformatics, Yale University, New Haven, Connecticut, 06520, USA}
\affiliation{Integrated Graduate Program in Physical and Engineering Biology, Yale University, New Haven, Connecticut, 06520, USA}
\affiliation{Department of Physics, Yale University, New Haven, Connecticut, 06520, USA}
\affiliation{Department of Applied Physics, Yale University, New Haven, Connecticut, 06520, USA}

\date{\today}

\begin{abstract}
Proteins fold to a specific functional conformation with a densely packed hydrophobic core that controls their stability. We develop a geometric, yet all-atom model for proteins that explains the universal core packing fraction of $\phi_c=0.55$ found in experimental measurements. We show that as the hydrophobic interactions increase relative to the temperature, a novel jamming transition occurs when the core packing fraction exceeds $\phi_c$. The model also recapitulates the global structure of proteins since it can accurately refold to native-like structures from partially unfolded states.
\end{abstract}

\keywords{Protein core packing, jamming, protein folding, glasses}

\maketitle

In native solution conditions, globular proteins fold from an extended chain to a compact, specific, and functional state. Protein folding is believed to be an equilibrium collapse process toward a global energy minimum driven primarily by the hydrophobicity of the amino acid sequence ~\cite{folding:DillBiochemistry1990,energylandscape:BryngelsonPNAS1987,energylandscape:LeopoldPNAS1992,energylandscape:WolynesScience1995,energylandscape:OnuchicAnnRevPhysChem1997,energylandscape:PlotkinQRevBiophys2002,folding:PaceJMB2011}. In addition, it is well known that proteins possess dense, solvent-inaccessible, or core, regions, which include $\sim 10\%$ of the protein and provide their thermal stability. Focusing on the hard-core atomic interactions, initial calculations of the core packing fraction found that $\phi \sim 0.7$-$0.74$, which is close to the maximum packing fraction in crystalline solids~\cite{packing:ChothiaNature1975,packing:RichardsJMB1974,packing:RichardsAnnRevBiophys1977,packing:LiangBPJ2001}.  In such hard-particle models, the maximum packing fraction corresponds to a minimum in the potential energy, i.e. $V \sim 1/\phi$. However, we now know that achieving such large values for the packing fraction is not possible for disordered states like protein cores, without causing interatomic overlaps. More recent work shows that the average packing fraction in globular protein cores is $\langle \phi \rangle = 0.55 \pm 0.01$~\cite{subgroup:GainesPRE2016}. 

While the fact that each protein folds to a specific conformation suggests an equilibrium process, dense packing in protein cores suggests that non-equilibrium processes also occur. For example, as hard particles are compressed, the system becomes rigid and solid-like, i.e. jammed, at a sufficiently large packing fraction $\phi_c$~\cite{jamming:OHernPRE2003}. However, the $\phi_c$ and mechanical properties depend on the protocol used to generate the particle packings, and thus jamming is a highly non-equilibrium process~\cite{jamming:ChaudhuriPRL2010,jamming:AshwinPRL2013,jamming:BertrandPRE2016,polydisperse:OzawaSciPost2017}. In addition, vibrational studies of proteins show that they posses a boson peak, or an abundance of low frequency modes in the density of states, which is a prominent feature of non-equilibrium systems, such as glasses~\cite{glassyproteins:PerticaroliSoftMatter2013,glassyproteins:PerticaroliBPJ2014,glassyprotein:NakagawaBPJ2019,glassyproteins:MoriPRE2020}. Moreover, recent experiments on the dry molten globule state~\cite{drymoltenglobule:BaldwinProteins2010,drymoltenglobule:JhaPNAS2009,drymoltenglobule:SarkarNPJ2013,drymoltenglobule:NeumaierJMB2014} suggest that the final stages of core formation take much longer than the initial stages of folding~\cite{drymoltenglobule:WilsonPRL2024}.  
Thus, it is important to develop a geomertic, yet atomistically accurate model for proteins, which will allow us to rigorously connect the nonequilibrium physics of hard-particle packings~\cite{jamming:OHernPRE2003,glasses:CharbonneauAnnRevCond2017} to the nearly folded conformational landscape of proteins~\cite{folding:DillScience2012}.

In this Letter, we first discuss a hard-sphere (HS) model for proteins with stereochemical constraints and a specific set of atom sizes that recapitulates the allowed backbone and side chain dihedral angle distributions in proteins. We then add attractive atomic interactions that scale with amino acid hydrophobicity (i.e. the HS+HP model) to explore core formation. We find that the HS+HP model collapses as the attractive strength relative to temperature is increased, and similar to jamming, transitions from a floppy to rigid state at $\phi_c \sim 0.55$. Additionally, we find that the potential energy of atomic overlaps scales as a power-law with packing fraction, $\langle V_r \rangle \sim (\phi - \phi_c)^{\delta}$ with a novel scaling exponent $\delta \sim 9/2$. This result suggests that proteins collapse until the core amino acids reach a mechanically stable state that resists the compression induced by the hydrophobic attractions. Moreover, HS+HP model proteins can refold from partially unfolded states, suggesting that the model can recapitulate the protein conformational landscape.

\begin{figure*}
\begin{center}
\includegraphics[width=\textwidth]{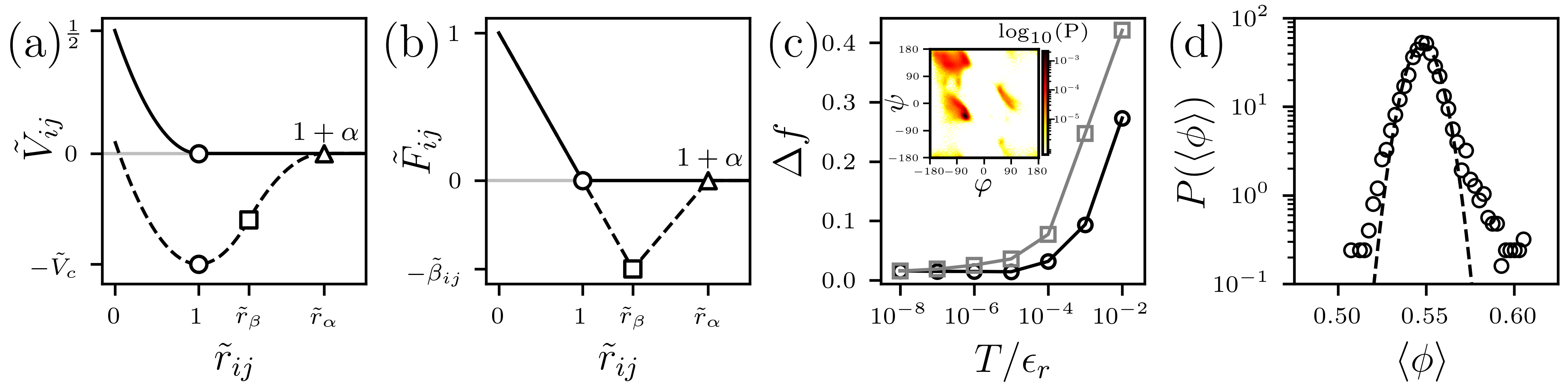}
\caption{The nonbonded dimensionless pair potential $\tilde{V}_{ij} = V_{ij} / \epsilon_r$ plotted versus atomic separation $\tilde{r}_{ij} = r_{ij} / \sigma_{ij}$ for purely-repulsive interactions (Eq.~\ref{eq:repulsive}) (solid line) and attractive interactions (Eq.~\ref{eq:ra}) (dashed line) and (b) the corresponding dimensionless force $\tilde{F}_{ij} = F_{ij}\sigma_{ij}/\epsilon_r$. The symbols represent the onset of repulsive interactions where $\tilde{r}_{ij} = 1$ and $\tilde{V}_{ij} = -\tilde{V}_c = -V_c / \epsilon_r$ (circles), the change in spring constant where $\tilde{r}_{ij} = \tilde{r}_{\beta} = 1 + r_{\beta}\sigma_{ij}$ and $\tilde{F}_{ij} = -\tilde{\beta}_{ij} = -\beta\lambda_{ij}\sigma_{ij}/\epsilon_r$ (squares), and the separation above which the interactions are zero $\tilde{r}_{ij} = \tilde{r}_{\alpha} = 1 + r_{\alpha}/\sigma_{ij}=1+\alpha$ (triangles). (c) The difference $\Delta f$ between the average fraction of backbone dihedral angle outliers (black circles) and side chain dihedral angle outliers (grey squares) between the HS model and proteins from a high-quality x-ray crystal structure database plotted versus the temperature $T/\epsilon_r$ at which the HS model proteins were simulated. Inset: The probability distribution of backbone dihedral angles $P(\varphi,\psi)$ sampled by high-quality x-ray crystal structures of proteins. The colors from light to dark indicate increasing probability on a logarithmic scale. (d) Probability distribution of the average core packing fraction $P(\langle \phi \rangle)$ in high-quality x-ray crystal structures of proteins calculated using the optimized HS atom sizes on a semi-log plot with a Gaussian fit (black dashed line).}
\label{fig:setup}
\end{center}
\end{figure*}

First, to calculate the core packing fraction, the set of atomic diameters $\{\sigma_i\}$ must be defined. However, the literature provides a wide range of possible $\{\sigma_i\}$ for the hard-core atomic interactions in proteins~\cite{subgroup:GainesPRE2016}. Therefore, we propose that $\{\sigma_i\}$ can be selected by validating the atom sizes against a fundamental feature of protein structure. Ramachandran, {\it et al.} first demonstrated that by assuming only repulsive, hard-core atomic interactions, plus the stereochemistry of amino acids, one can predict the backbone dihedral angle pairs $\varphi$ and $\psi$ that occur in proteins are those pairs that do not cause large atomic overlaps~\cite{rama:RamachandranJMB1963,rama:RamakrishnanBPJ1965}. (See the inset of Fig.~\ref{fig:setup}(c) for the backbone dihedral angle distribution from high-quality x-ray crystal structures of proteins.) We have also validated this approach for the distributions of side chain dihedral angles~\cite{subgroup:ZhouBPJ2012,subgroup:ZhouProteins2014,subgroup:CaballeroProtSci2014,subgroup:CaballeroPEDS2016,subgroup:GainesPEDS2017}. The potential energy for the HS model includes both nonbonded and bonded atomic interactions. For the nonbonded interactions, we employ a purely repulsive linear spring potential to prevent atomic overlaps,  
\begin{equation}
    \frac{V_r(r_{ij})}{\epsilon_r} = \frac{1}{2} \left( 1 - \frac{r_{ij}}{\sigma_{ij}} \right)^2 \Theta \left(1-\frac{r_{ij}}{\sigma_{ij}} \right),
\label{eq:repulsive}
\end{equation}
where $\epsilon_r$ defines the repulsive energy scale, $r_{ij}$ is the center-to-center distance between atoms $i$ and $j$, $\sigma_{ij}$ is their average diameter, and $\Theta(x)$ is the Heaviside step-function. (See Fig.~\ref{fig:setup} (a) and (b).) The total repulsive potential energy includes all atom pairs except those that participate in bonded interactions: $V_r = \sum_{{\langle i, j \rangle}'} V_r(r_{ij})$, where $\langle i, j\rangle'$ is the set of nonbonded atom pairs. We add restraints on the bond lengths $r_{ij}$, bond angles $\theta_{ijk}$, and dihedral angles $\omega_{ijkl}$ with rest values, $r_{ij}^0$, $\theta_{ijk}^0$, and $\omega_{ijkl}^0$ that occur in each target protein's high-resolution x-ray crystal structure:
\begin{equation}
\label{eq:bonded}
\frac{V_{b}(r_{ij})}{k_b} = \frac{1}{2\sigma_H^2} \left( r_{ij} - r_{ij}^{0} \right)^2,
\end{equation}

\begin{equation}
\label{eq:angle}
\frac{V_{a}(\theta_{ijk})}{k_a} = \frac{1}{2} \left( \theta_{ijk} - \theta^0_{ijk} \right)^2,
\end{equation}

\begin{equation}
\label{eq:dihedral}
\frac{V_{d}(\omega_{ijkl})}{k_d} = \frac{1}{2} \left( \omega_{ijkl} - \omega_{ijkl}^{0} \right)^2,
\end{equation}
where $k_b = k_a = k_d= \epsilon_r$ are the respective spring constants and $\sigma_H$ is the diameter of hydrogen. (Below, all energy scales will be given in units of $\epsilon_r$.) We set the spring constants to be equal to weight nonbonded overlaps and deformations in stereochemistry equally. Additionally, only the dihedral angles $\omega_{ijkl}$ needed to maintain high-quality protein stereochemistry are restrained. First, we add restraints to the main chain peptide bond dihedral angle $\omega_{ijkl}$, which due to the peptide bond's partial double-bonded character, is relatively planar in high-quality protein structures. Second, amino acids with side chains containing double bonds require restraints to maintain their planar geometry, such as in the phenylalanine ring. The total potential energy for the HS model is then $V = V_r + V_b + V_a + V_d$, where $V_b = \sum_{\langle i, j \rangle} V_b(r_{ij})$, $V_a = \sum_{\langle i, j, k \rangle} V_a(\theta_{ijk})$, $V_d = \sum_{\langle i, j, k, l \rangle} V_d(\omega_{ijkl})$, $\langle i, j\rangle$ is the set of bonded atom pairs, $\langle i, j, k\rangle$ is the set of atom triples that defines each bond angle, and $\langle i, j, k, l\rangle$ is the set of groups of four atoms that define the dihedral angles. All hydrogens are placed using the \textsc{Reduce} software~\cite{reduce:WordJMB1999}. When comparing simulation results to experimentally determined protein structures, we use a high-resolution dataset of $\sim 5,000$ structures with a resolution $< 1.8~$\AA~ culled from the Protein Data Bank (PDB)~\cite{PISCES:WangBioinformatics2003,PISCES:WangNucleicAcids2005}. For the HS protein simulations, we carry out Langevin dynamics over range of temperatures $10^{-8} < T/\epsilon_r < 10^{-2}$ using $20$ randomly selected, single chain target proteins with no disulfide bonds from the x-ray crystal structure dataset. Sizes range from $N_{\rm{aa}}=60$-$335$ with an average of $\langle N_{\rm{aa}} \rangle=150$ amino acids. (PDBIDs are given in Table S1 and examples of the restraints in Eqs.~\ref{eq:bonded}-\ref{eq:dihedral} are given in Tables S2 and S3 in Supplemental Material (SM).) 

\begin{figure*}
\begin{center}
\includegraphics[width=\textwidth]{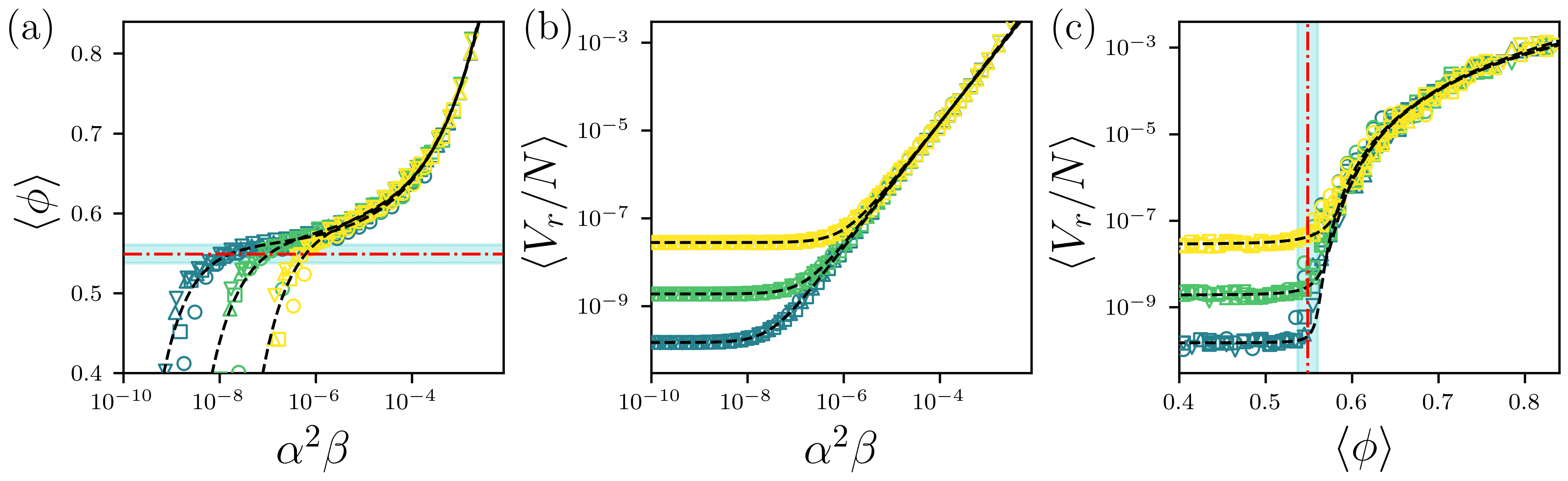}
\caption{(a) The average core packing fraction $\langle \phi \rangle$ plotted versus the attraction strength $\alpha^2\beta$ for the HS+HP protein model for temperatures $T/\epsilon_r = 10^{-6}$ (yellow), $10^{-7}$ (green), and $10^{-8}$ (blue) and $\alpha = 0.5$ (circles), $1.0$ (squares), $1.5$ (upward triangles), and $2.0$ (downward triangles). The horizontal red dot-dashed line and cyan shading indicate the average and standard deviation of the core packing fraction in the high-resolution x-ray crystal structure data set. The black dashed lines indicate fits to Eq.~\ref{eq:double_power}. (b) The average repulsive potential energy per atom  $\langle V_r/N \rangle$ plotted versus $\alpha^2\beta$. The black dashed lines indicate fits to Eq.~\ref{eq:power_plat}. (c) $\langle V_r/N \rangle$ plotted versus $\langle \phi \rangle$. The vertical red dot-dashed line and cyan shading indicate the average and standard deviation of the core packing fraction in the high-resolution x-ray crystal structure data set. The black dashed lines indicate fits to Eq.~\ref{eq:combo}.}
\label{fig:xtal_jamming}
\end{center}
\end{figure*}

Using optimized $\{\sigma_i\}$ (Table S4 of SM)~\cite{subgroup:ZhouBPJ2012,subgroup:ZhouProteins2014,subgroup:CaballeroProtSci2014,subgroup:CaballeroProteins2015,subgroup:CaballeroPEDS2016,subgroup:GainesPEDS2017}, we compare the backbone and side chain dihedral angles sampled by the HS model and those of high-quality x-ray crystal structures. We use the software package \textsc{MolProbity} to quantify the fraction $f$ of backbone and side chain dihedral angle outliers, with respect to a reference set of high quality x-ray crystal structures~\cite{molprob:DavisNAR2007,molprob:ChenACSD2010,molprob:WilliamsProtSci2018}. We compare the fraction of backbone and side chain outliers in the HS simulations $f_s(T/\epsilon_r)$ to the fraction of outliers in our high-resolution x-ray crystal structure database $f_x$. We show in Fig.~\ref{fig:setup} (c) that $\Delta f(T/\epsilon_r) = f_s(T/\epsilon_r) - f_x$ approaches zero for both backbone and side chain dihedral angles as $T/\epsilon_r$ decreases (and the HS model approaches the hard-core limit). Note that the HS protein model recapitulates the Ramachandran map even though it has fewer restraints than in typical all-atom protein force fields. For example, in an alanine dipeptide, the HS model includes two dihedral angle restraints, whereas current Amber and CHARMM force fields have $41$ dihedral angle restraints~\cite{a99sb-ildn:Lindorff-LarsenProteins2010,charmm36m:HuangNatMet2017}.

With this optimized set of atomic diameters $\{\sigma_i\}$, we can calculate the average core packing fraction $\langle \phi \rangle$ in the high-resolution x-ray crystal structure data set as shown in Fig.~\ref{fig:setup} (d). Core amino acids are those that have relative solvent accessible surface area rSASA $< 10^{-3}$, using the Lee and Richards algorithm with a probe size of a water molecule~\cite{FreeSASA:Mitternacht2016}. As we have previously reported~\cite{subgroup:GainesPRE2016,subgroup:MeiProteins2020}, we find $\langle \phi \rangle \sim 0.55 \pm 0.01$. The same result is found for solution NMR structures when only including high quality bundles~\cite{subgroup:GrigasProSci2022}. An important question naturally arises, why does the folding process give rise to this value for $\langle \phi \rangle$ in all globular protein cores?

To study core formation, we can add attractive interactions to the HS protein model, which yields the HS+HP model. For the nonbonded attractive interactions between atoms, we extend the potential in Eq.~\ref{eq:repulsive} to $r_{\beta} / \sigma_{ij} = 1 + \sigma_{ij} \beta_{ij}/\sigma_H$ and cutoff the interactions at $r_{\alpha}/\sigma_{ij} = 1+\alpha > r_{\beta}$ using piecewise harmonic functions of $r_{ij}$: 
\begin{equation}
\label{eq:ra}
\frac{V_{a}(r_{ij})}{\epsilon_r} = \begin{cases}
    \frac{1}{2} \left( 1 - \frac{r_{ij}}{\sigma_{ij}} \right)^2 - V_c/\epsilon_r ~~{\rm for}~r_{ij} \leq r_{\beta} \\
    -\frac{k}{2\epsilon_r} \left( \frac{r_{ij}}{r_{\alpha}} - 1 \right)^2 ~~~~~{\rm for}~ r_{\beta} < r_{ij} \leq r_{\alpha} \\
    0 ~~~~~~~~~~~~~~~~~~~~~~~~~~~~{\rm for}~ r_{ij} > r_{\alpha},
\end{cases}
\end{equation}
where $V_c/\epsilon_r =  (k/\epsilon_r)\left( r_{\beta}/r_{\alpha} - 1 \right)^2/2  + \left( 1 - r_{\beta}/\sigma_{ij}\right)^2/2$ for continuity. $\alpha$ defines the attractive range and $\beta_{ij}= \beta \lambda_{ij}$ defines the magnitude of the attractive force. (See Fig.~\ref{fig:setup} (a) and (b).) $\lambda_{ij} = (\lambda_i+\lambda_j)/2$ is the average hydrophobicity associated with atom pairs $i$ and $j$, where $1 \le \lambda_i \le 0$ is the hydrophobicity per amino acid and is assigned to each atom on a given amino acid~\cite{subgroup:SmithPRE2014}. (See Table S5 in SM.)

To explore the dynamics of folding for the HS+HP model, we run Langevin dynamics with the HS-energy minimized x-ray crystal structure of a given protein as the initial condition. We consider $20$ randomly selected single-chain protein targets from the high-resolution x-ray crystal structure database and study the folowing parameter regimes: $0.5 \le \alpha \le 2$, $10^{-12} \le \beta \le 10^{-3}$, and $10^{-8} \le T/\epsilon_r \le 10^{-6}$.  In Fig.~\ref{fig:xtal_jamming} (a), we show the packing fraction of core residues $\langle \phi\rangle$ averaged over the $20$ proteins versus increasing attractive strength, quantified using $\alpha^2 \beta$. Plotting $\langle \phi \rangle$ versus $\alpha^2 \beta$ collapses the data for each temperature $T/\epsilon_r$. At small $\alpha^2\beta$, the proteins unfold and $\langle \phi \rangle < 0.55$. As the attractive interactions increase, a plateau at $\langle \phi \rangle \sim 0.55$ (i.e. at the average packing fraction of experimentally determined protein cores) occurs for $\alpha^2\beta \sim T/\epsilon_r$. Increasing the attraction further causes a steep increase in $\langle \phi \rangle$. As $T/\epsilon_r$ is lowered, the HS+HP model behaves as a hard-core system and the plateau extends to smaller $\alpha^2 \beta$. $\langle \phi \rangle$ versus $\alpha^2 \beta$ is well fit by
\begin{equation}
    \langle \phi \rangle = A \left( \alpha^2\beta  \right)^a -  B \left( \alpha^2\beta  \right)^{-b} + \phi_c,
\label{eq:double_power}
\end{equation}
where $A$ and $B$ are constants, $\phi_c \rightarrow 0.55$ and the exponents $a \rightarrow 1/3$ and $b \rightarrow 2$ as $T/\epsilon_r \rightarrow 0$. 

How are such large values of $\langle \phi \rangle > 0.55$ possible in Fig.~\ref{fig:xtal_jamming} (a)? When we plot the average total nonbonded repulsive potential energy per atom $\langle V_{r}/N \rangle$ versus $\alpha^2\beta$ in Fig.~\ref{fig:xtal_jamming} (b), we find that $\langle V_r / N \rangle \sim V_0$, where $V_0 \sim T/\epsilon_r$ for $\alpha^2\beta < T/\epsilon_r$. However, when $\alpha^2\beta > T/\epsilon_r$, $\langle V_r/N \rangle$ increases from the plateau value $V_0$ as a power-law:
\begin{equation}
    \langle V_r / N \rangle -V_0= C \left( \alpha^2\beta \right)^c,
\label{eq:power_plat}
\end{equation}
where $C$ is a constant and $c \rightarrow 3/2$ as $T/\epsilon_r \rightarrow 0$.
Thus, we find that when $\langle \phi \rangle > 0.55$, the total repulsive energy per atom increases strongly, which indicates a jamming transition.

In Fig.~\ref{fig:xtal_jamming} (c), we combine data from Figs.~\ref{fig:xtal_jamming} (a) and (b). For $\langle \phi \rangle < 0.55$, $\langle V_{r} / N \rangle \sim V_0$. When $\langle \phi \rangle > 0.55$, $\langle V_r/N \rangle$ increases as a power-law, obtained by combining Eqs.~\ref{eq:double_power} and~\ref{eq:power_plat}:
\begin{equation}
    \langle \phi \rangle = \mathcal{A} (\Delta V_r)^{a/c} + \mathcal{B} (\Delta V_r)^{-b/c} +  \phi_c,
\label{eq:combo}
\end{equation}
where $\Delta V_r =  \langle V_r / N \rangle - V_0$, $\mathcal{A} = A / C^{a/c}$ and $\mathcal{B} = B / C^{-b/c}$.  When $\langle \phi \rangle - \phi_c = \langle \Delta \phi \rangle \gg 0$, Eq.~\ref{eq:combo} simplifies to $\langle V_r / N \rangle \sim \langle \Delta \phi \rangle^\delta$, where $\delta = c/a \rightarrow 9/2$ in the $T/\epsilon_r \rightarrow 0$ limit. This result is similar to that found for the jamming transition in particle packings, except with a significantly larger exponent than $\delta = 2$ expected from affine compression.

Thus, Fig.~\ref{fig:xtal_jamming} shows that the HS+HP model undergoes a jamming transition when the average packing fraction increases above the value observed in x-ray crystal structures of proteins. However, the jamming transition in the HS+HP model has a scaling exponent $\delta$ that is more than a factor of two larger than that found previously for hard-sphere systems and bead-spring polymers~\cite{subgroup:GrigasPRE2024}. In the SM, we confirm that the collapse transition in bead-spring polymers with the same nonbonded interactions and only bond-length constraints yields $\delta=2$, which suggests that the anomalous exponent for the HS+HP model is caused by the unique geometry of amino acids and not from the attractive interactions. A possible source of the anomalous scaling exponent is changes in the number of contacts between amino acids as the core is compressed~\cite{subgroup:VanderWerfPRE2020}. 

\begin{figure}
\begin{center}
\includegraphics[width=\columnwidth]{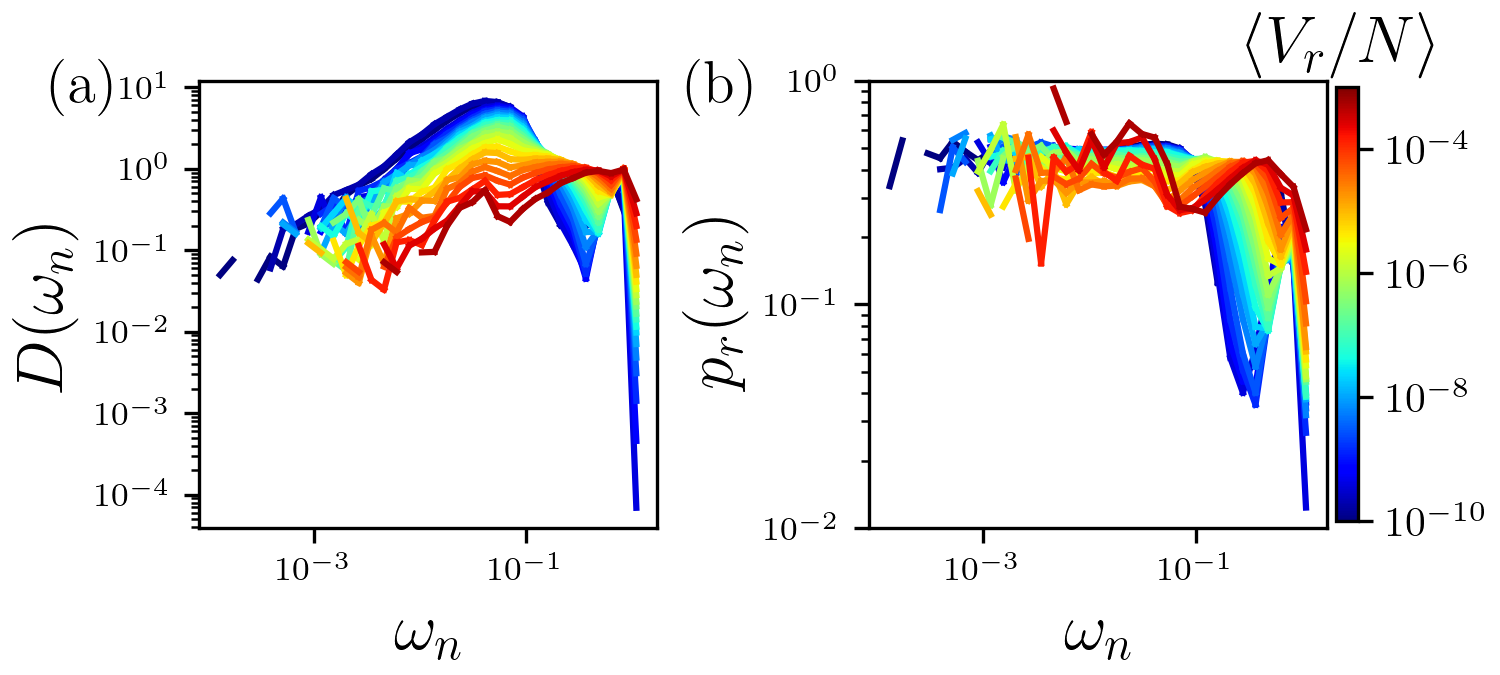}
\caption{(a) The vibrational density of states (VDOS) $D(\omega_n)$, where $\omega_n$ is the frequency, of the C$_{\alpha}$ atoms in the HS+HP model at $T/\epsilon_r = 10^{-8}$ for all $\alpha$ and $\beta$ in Fig.~\ref{fig:xtal_jamming}. (b) Participation ratio $p_r(\omega_n)$ plotted versus $\omega_n$. The average total nonbonded repulsive potential energy per atom $\langle V_r/N \rangle$ increases from blue to red on a logarithmic scale.}
\label{fig:VDOS}
\end{center}
\end{figure}

Below jamming onset, unjammed systems possess a large number of low frequency, liquid-like modes in the vibrational density of states (VDOS). Near jamming onset, excess intermediate frequencies, known as the boson peak, occur in the VDOS, and as the packing fraction increases above jamming onset the boson peak is suppressed~\cite{vdos:SilbertPRL2005,vdos:WyartPRE2005}. We calculate the VDOS from the eigenvalues $e_n$ of the displacement correlation matrix $S = VC^{-1}$, where $V_{ij} = \langle v_i v_j \rangle$ is the velocity correlation matrix and $C_{ij} = \langle (r_i - r_i^0) (r_j - r_j^0) \rangle$ is the positional covariance matrix, $v_i$ are the atom velocities, $r_i$ are the atom positions, and $r_i^0$ are the average atom positions. The angle brackets indicate time averages. Each eigenvalue $e_n$ has a corresponding eigenvector ${\hat e}_n = \{e_{1xn},e_{1yn}, e_{1zn},\ldots,e_{Nxn},e_{Nyn},e_{Nzn} \}$. The VDOS $D(\omega_n)$ is then obtained by binning the frequencies $\omega_n = \sqrt{e_n}$~\cite{jamming:HenkesSoftMatter2012,jamming:BertrandPRE2014}, where the frequencies are given in units of $\epsilon_r/\sqrt{m_H \sigma_H^2}$, where $m_H$ is the mass of hydrogen. 

In Fig.~\ref{fig:VDOS} (a), to investigate the rigidification of the HS+HP model, we plot the VDOS of the backbone C$_{\alpha}$ atoms averaged over the $20$ proteins for $T/\epsilon_r = 10^{-8}$. We show $D(\omega_n)$ for all $\alpha$ and $\beta$ values in Fig.~\ref{fig:xtal_jamming} and as a function of $\langle V_r / N \rangle$ to identify the jamming transition. When the HS+HP proteins are unjammed with $\langle V_r/N \rangle \sim 10^{-10}$, the VDOS possesses a large peak of liquid-like modes in the range $10^{-2} < \omega_n < 10^{-1}$, as well as a secondary peak near $\omega_n \sim 1$ corresponding to the bonded interactions. As $\langle V_r/N \rangle$ increases, the liquid-like peak decreases and the modes at intermediate frequencies fill-in to form a plateau near $D(\omega_n) \sim 1$. A key sign of the onset of rigidification is the formation of a plateau in the intermediate frequency region of the VDOS, also known as the boson peak~\cite{jamming:LiuAnnRevConMatPhys2010}. The boson peak is suppressed when the system becomes overcompressed with increasing $\langle V_r /N\rangle$.  (The results for the VDOS of attractive bead-spring polymers are similar, except the liquid-like modes vanish more rapidly with increasing $\langle \phi \rangle$ related to the difference in $\delta$, as shown in SM.)

Near jamming onset in packings of spherical particles, the vibrational modes in the VDOS plateau region are quasi-localized, i.e. many particles participate in the eigenmodes, but they are not phonon-like~\cite{jamming:LiuAnnRevConMatPhys2010}. We investigate the localization of the modes in the platueau region of $D(\omega_n)$ by calculating the participation ratio for each eigenmode,
\begin{equation}
    p_r(\omega_n) = \frac{1}{N} \frac{\left( \sum_{i=1}^{N} | \vec{d}_i(\omega_n) |^2 \right)^2}{\sum_{i=1}^{N} | \vec{d}_i(\omega_n) |^4 },
\end{equation}
where ${\vec d}_i(\omega_n) = \{ e_{ixn},e_{iyn}, e_{izn} \}$ is the contribution of particle $i$ to the $n$th eigenvector of $S$~\cite{vdos:ArceriPRL2020}. In Fig.~\ref{fig:VDOS} (b), we plot the binned $p_r(\omega_n)$ at each $\langle V_r/N \rangle$. A key difference between $p_r(\omega_n)$ for the unjammed systems ($\langle V_r/N\rangle \sim 10^{-10}$) and jammed systems is that there is a strong increase in $p_r(\omega_n)$ for frequencies in the range $10^{-1} < \omega_n < 10^{0}$, which indicates the development of quasi-localized modes at intermediate frequencies. As expected, the highest frequencies correspond to local excitations.

\begin{figure}
\begin{center}
\includegraphics[width=\columnwidth]{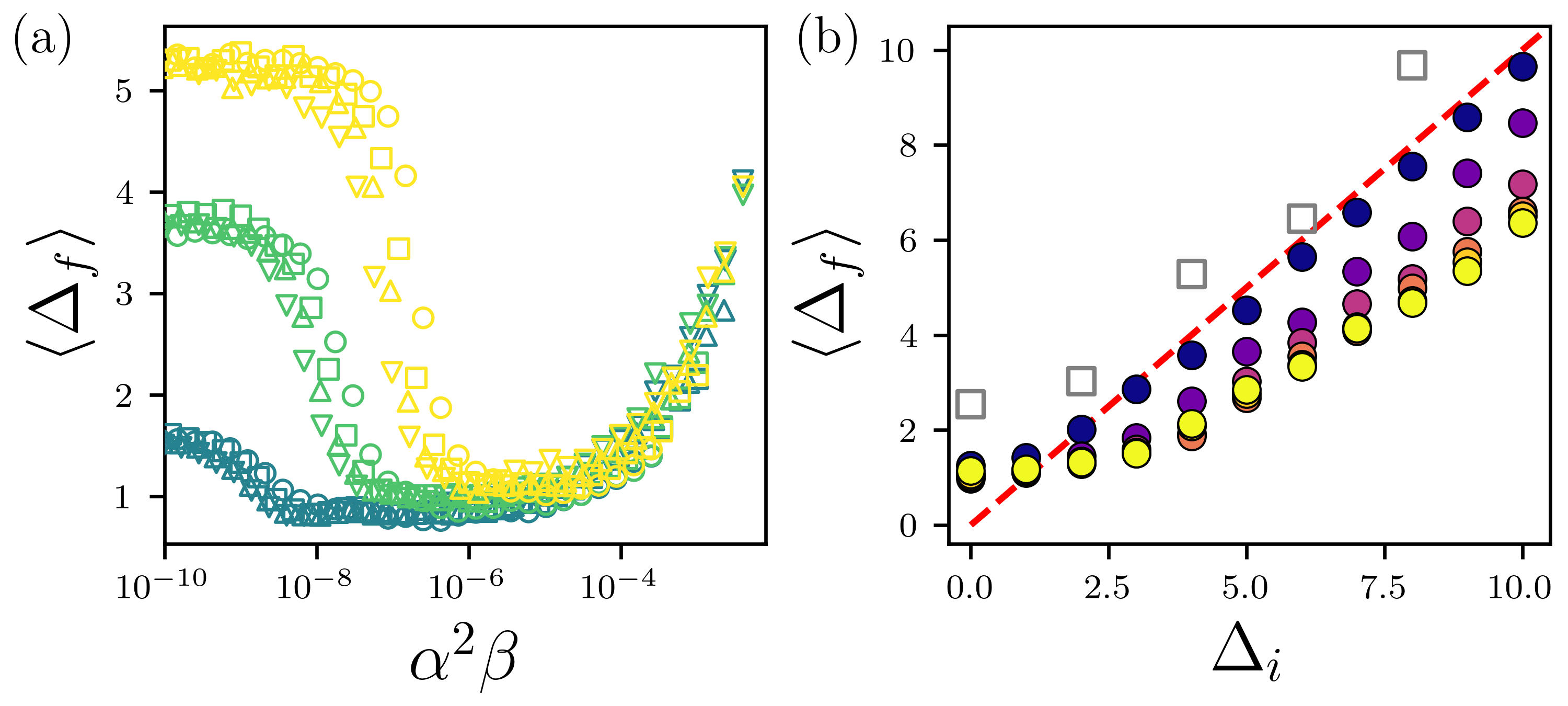}
\caption{(a) C$_\alpha$ RMSD $\langle \Delta_f\rangle$ in \AA~between the HS+HP model proteins and the x-ray crystal structures averaged over $20$ proteins plotted versus $\alpha^2\beta$ when starting from the experimental structure for temperature $T/\epsilon_r = 10^{-6}$ (yellow), $10^{-7}$ (green), and $10^{-8}$ (blue) and $\alpha = 0.5$ (circles), $1.0$ (squares), $1.5$ (upward triangles), and $2.0$ (downward triangles). (b) Average C$_\alpha$ RMSD $\langle \Delta_f \rangle$ plotted versus the initial C$_{\alpha}$ RMSD $\Delta_i$ in \AA~for $T/\epsilon_r = 10^{-7}$. The filled circles are colored by $\alpha=0.5-5.5$ increasing from purple to yellow, and $\beta$ is set so that $\alpha^2\beta \sim T/\epsilon_r$. All-atom MD simulations of a single protein (PDBID: 2IGP) using the Amber99SB-ILDN force field are shown as grey squares. The red dashed line indicates $\langle \Delta_f \rangle = \Delta_i$.}
\label{fig:rmsd}
\end{center}
\end{figure}

We demonstrated that during folding, the HS+HP model for proteins undergoes a jamming transition at the average core packing fraction observed in high-resolution x-ray crystal structures. We now quantify whether the backbone atoms of the HS+HP model deviate from the x-ray crystal structures during the jamming process.  To do this, we calculate the root-mean-square-deviations (RMSD) in the C$_{\alpha}$ positions between the simulated and experimental protein structures,
\begin{equation}
    \Delta = \sqrt{\frac{1}{N_{\rm{aa}}} \sum_{m=1}^{N_{\rm aa}} ( {\vec r}_{ms} - {\vec r}_{me} )^2 },
\end{equation}
where ${\vec r}_{ms}$ and ${\vec r}_{me}$ are the C$_{\alpha}$ positions of the $m$th amino acid from the simulations and x-ray crystal structures, respectively. We find that $\Delta$ converges rapidly as a function of time, and thus we focus on $\Delta_f$ at the last time point in the simulations. We plot $\langle \Delta_f \rangle$ averaged over the $20$ proteins in Fig.~\ref{fig:rmsd} (a) for the HS+HP simulations presented above. We find that $\langle \Delta_f \rangle \sim 1$~\AA~near jamming onset, confirming that not only the core packing fraction, but also the overall backbone conformation is similar to the x-ray crystal structure near jamming onset.

Does the C$_{\alpha}$ RMSD of the HS+HP model relative to the x-ray crystal structures remain small when the simulations are initialized further from the x-ray crystal structure? To study the ability of the HS+HP model to refold proteins, we initialize the HS+HP simulations with conformations at different values of the C$_{\alpha}$ RMSD $\Delta_i$ using the HS model conformations, which unfold over time since there are no attractive forces.  We then run Langevin dynamics simulations of the HS+HP model at $T / \epsilon_r = 10^{-7}$ over the range $0.5 \le \alpha \le 5.5$ and we set $\beta$ such that $\alpha^2 \beta \sim T/\epsilon_r$. 

In Fig.~\ref{fig:rmsd} (b), we plot the long-time C$_\alpha$ RMSD $\langle \Delta_f \rangle$ versus $\Delta_i$ for a range of $\alpha$ averaged over all $20$ proteins studied. We find that for short attractive ranges (i.e. $\alpha \lesssim 0.5$), while when starting in the crystal structure can lead to a jamming transition, the HS+HP model cannot refold (i.e. $\langle \Delta_f \rangle \sim \Delta_i$) above $\Delta_i \sim 2$~\AA. As $\alpha$ is increased, the HS+HP model can refold initial states with $\Delta_i \lesssim 5$ \AA~to $\langle \Delta_f \rangle \sim 2$ \AA, a threshold that is considered properly folded in all-atom MD simulations of protein folding~\cite{folding:Lindorff-LarsenScience2011}. In addition, all HS+HP proteins that refold to form a well-defined core possess $\langle \phi \rangle \sim 0.55$. We also compared these results to those from all-atom MD simulations using the Amber99SB-ILDN force field in explicit water. We find that $\langle \Delta_f \rangle \sim 2$~\AA~when starting near the x-ray crystal structure for PDBID: 2IGP, yet $\langle \Delta_f \rangle \sim \Delta_i$ when $\Delta_i >2$~\AA~after running for $> 1~\mu$s~\cite{Amber:BestJPCB2009,a99sb-ildn:Lindorff-LarsenProteins2010,gromacs:AbrahamSoftwareX2015,tip3p:JACS1981,tip3p:MarkJPCA2001,vrescale:BussiJCP2007}. (More details are found in SM.)

Here, we have developed a quantitatively accurate model for protein structure in which the stereochemistry of the amino acids is preserved and the atom sizes are optimized to recapitulate the experimentally observed backbone (and side chain) dihedral angle distributions. By adding hydrophobic attractive interactions, we showed that a novel jamming transition occurs during folding at the average core packing fraction of protein x-ray crystal structures. We showed that the total repulsive potential energy versus $\langle \Delta \phi \rangle$ obeys power-law scaling above jamming onset with an anomalous exponent $\delta > 2$. In addition, the vibrational response indicates that the HS+HP model rigidifies at $\phi_c = 0.55$ with quasi-localized vibrational modes at intermediate frequencies. Thus, we have demonstrated that the core packing fraction observed in high quality experimental protein structures is due to the onset of jamming under hydrophobic compression and have provided a theoretical direction for understanding non-equilibrium properties of proteins. In addition, starting from partially unfolded states with $\Delta_i \lesssim 5$~\AA, HS+HP proteins can refold to the x-ray crystal structure. We believe that the HS+HP model is well-suited for tackling many open problems in protein science, such as predicting the structural response to amino acid mutations, identifying protein-protein interactions, and understanding protein structure {\it in vivo}~\cite{mutation:XuProtSci1998,mutation:LiuJMB2000,mutation:BaaseProtSci2010,binding:ChenProtSci2013,invivo:DavisCurOpinStructBio2018}. In addition, the HS+HP model can be used to investigate the effects of folding rate on protein core packing, given that the properties of other jammed systems possess strong cooling rate dependence~\cite{aging:HuNatPhys2016,aging:LiChemSci2022,glass:LiaoPNAS2023,glassyprotein:DorbathJCP2024}. 

\begin{acknowledgments}
The authors acknowledge support from NIH Training Grant No. T32GM145452 (A. T. G. and C. S. O.), NIH Training Grant No. T15LM007056-37 (J. A. L.) and the High Performance Computing facilities operated by Yale’s Center for Research Computing.
\end{acknowledgments}

\end{document}